\documentstyle[preprint,prc,aps]{revtex}

\newcommand{\gea}{\raisebox{-.3ex}{\small $ \
\stackrel{\textstyle >}{\sim} $ }}
\newcommand{\be}{\begin{equation}}
\newcommand{\ee}{\end{equation}}

\begin{document}

\draft
\title{
Skyrme-model $\pi NN$ form factor and nucleon-nucleon
interaction}
\author{G. Holzwarth\footnote{Supported by the Volkswagen-Stiftung}}
\address{Physics Department, Siegen University, 57068 Siegen, Germany}
\author{R. Machleidt}
\address{Physics Department, University of Idaho, Moscow, Idaho 83843}
\date{\today}

\maketitle

\begin{abstract}
We apply the strong $\pi NN$ form factor, which emerges from the Skyrme
model, in the two-nucleon system using a one-boson-exchange (OBE) model
for the nucleon-nucleon (NN) interaction.
Deuteron properties and phase parameters of NN scattering are reproduced
well. 
In contrast to the form factor of monopole shape that is traditionally
used in OBE models,
the Skyrme form factor leaves low momentum transfers essentially
unaffected while it suppresses the high-momentum region strongly.
It turns out that this behavior is very appropriate for models
of the NN interaction and makes possible to use a 
soft pion form factor in the NN system.
As a consequence, 
the $\pi N$ and the $NN$ systems can be described using the same
$\pi NN$ form factor, which is impossible with the monopole.
\end{abstract}


\section{Introduction}
It is well established that boson-exchange models are very successful
in describing the low-energy nucleon-nucleon (NN) interaction
\cite{Mac89}.
Examples for such models are the Nijmegen~\cite{NRS78},
Paris~\cite{Lac80}, and Bonn~\cite{MHE87} potentials~\cite{ML94}.
Typically, these models take into account the non-strange mesons with masses
below 1 GeV plus a $2\pi$-exchange contribution. 
If the latter is approximated
by a scalar-isoscalar boson (with mass 500--700 MeV),
one speaks of the
one-boson-exchange (OBE) model.

In meson-exchange models for the NN interaction, the
meson-nucleon vertices are, in general, multiplied with so-called form factors,
which are needed to avoid divergences in loop integrals.
While the vertices are derived from effective meson-nucleon
Langrangians which the models are based upon, the form factors
are introduced essentially {\it ad hoc} and do not emerge
from the underlying Lagrangians.
Though the substructure of hadrons provides, in principal, a
physical picture and justification for the form factors,
in most OBE models no attempt is made 
to use form factors
that have a theoretical basis in QCD or QCD-related models.
Instead, a phenomenological {\it ansatz} is used for
the form factor, like
\be
F_{\alpha}({\bf q}^2) = \left( \frac{\Lambda_\alpha^2-m_\alpha^2}
                            {\Lambda_\alpha^2+{\bf q}^2} \right)^{n_\alpha}
\; ,
\ee
where ${\bf q}$ is the three-momentum
transfer, $m_\alpha$ the mass of the exchanged meson,
and $\Lambda_\alpha$ the so-called cutoff mass;
$n_\alpha=1$ defines the monopole
form factor and $n_\alpha=2$ the dipole.
In the contruction of OBE potentials, the cutoff parameters
$\Lambda_\alpha$ are adjusted (together with the
meson-nucleon coupling constants) such as to yield an optimal fit
of the NN data. Typical values for $\Lambda_\alpha$
range between 1.3 and 2 GeV~\cite{MHE87}.

An example 
for a QCD-inspired form factor
is the cloudy-bag form factor 
for the pion~\cite{Tho83}, which is given by
\be
F_{CB}({\bf q}^2) = \frac{3 j_1(|{\bf q}|R)}{|{\bf q}|R} \; ,
\ee
where $j_1$ denotes the spherical Bessel function
and $R$ is the bag radius.
The cutoff mass used in Eq.~(1) with $n_\pi=1$ and  the $R$ used
in Eq.~(2) are roughly related by
$\Lambda_\pi = \sqrt{10}/R$, which implies $\Lambda_\pi \approx 780$
MeV for $R\approx 0.8$~fm.
Unfortunately, pion form factors with these 
(seemingly very reasonable)
parameters fail
in the NN system, since they cut out too much of the tensor force
provided by the pion: the deuteron quadrupole moment and asymptotic
D/S state ratio and the $\epsilon_1$ mixing parameter of NN scattering
(which all depend crucially on the nuclear tensor force) come out too 
small~\cite{note1}.
A possible cure for this problem is the introduction of new short-range
tensor-force generating mechanisms in the NN system, like 
the exchange of a heavy pion,
$\pi'(1300)$~\cite{PDG94}, 
which can also be viewed as a contribution from correlated
$\pi$-$\rho$ exchange~\cite{HT90}. However, 
this requires to take the meson-exchange
mechanism seriously at a very short distance between the interacting
nucleons (namely, a distance equivalent to an exchanged mass
of about 1300 MeV, that is $\approx 0.15$ fm). 
This may be in conflict with the implications
of a soft pion form factor ($R\approx 0.8$ fm), 
which leaves no room for the
exchange of mesons or meson systems heavier than 1 GeV.

Another aspect of the problem is that models for $\pi N$ 
scattering seem to require
a soft $\pi NN$ form factor ($\Lambda_\pi \approx 800$ MeV
or $R\approx 0.8$ fm),
if the analytic expressions Eqs.~(1) or (2) are used
for the $\pi NN$ form factor~\cite{TTM81}. 
Thus, with these types of form factors,
it is impossible to describe 
the $\pi N$ and $NN$ systems
consistently.

One reason for this problem may simply be that the shapes 
of the form factors conventionally used are not very appropriate.
Note that simplicity and convenience is traditionally
the main argument for Eq.~(1).

Recently the strong $\pi NN$ form factor has been extracted from
Skyrme type models which comprise the essential low-energy features of
QCD in effective nonlinear meson dynamics and the description of
nucleons as solitons in meson fields. It turned out that the shape of
the resulting form factor is quite different from the conventional
monopole form.
This suggests to take another look at the $NN$ system to find out whether
the implications of these models could be helpful for the form factor
discussion. 

The Skyrme model in its 'adiabatic' approximation 
is able to give a quite convincing and unified 
description of the essential features of the $\pi N$
system throughout and even beyond the resonance regions in all elastic
scattering channels except for the S and P channels~\cite{HEH84}. 
This qualitative statement should be seen in the light of the
fact that the model contains only {\it one} free parameter, the 
strength of the Skyrme term. For the S and P waves the adiabatic
approximation is not sufficient, due to the interplay between the
collective zero modes and the continuum of soliton fluctuations.
Although this makes it technically quite involved to analyse elastic
$\pi N$ scattering at low momentum transfers 
in the S and P channels, the Skyrme model has
been shown to provide the right amount of isospin-independent
background scattering and isospin splitting in the
S-channels~\cite{WMW92}, as well
as in the P13 and P31 channels~\cite{HPJ90}, 
and an accurate description of the P33
resonance~\cite{H93}. Only in the P11 channel 
the rise in the phase shift sets in at too low
energies due to the rather low-lying Roper resonance~\cite{HPJ90}. 
Again, this qualitative result is achieved with one parameter.
Extensions of the Skyrme model (to chiral order six, or inclusion of
vector mesons~\cite{SWHH89}) can improve 
the agreement in some instances at the
expense of additional parameters, but there has never been an attempt
to find an optimal version which would quantitatively cover the
experimental data in all scattering channels. 

It should, perhaps, be noted that these results for S and P waves at low
energies were obtained in a K-matrix unitarization which probably is
not very sensitive to the high-energy cutoff of an underlying form factor. 
But it is at very low momentum transfers where the Skyrme model
form factor deviates crucially from the standard monopole type, and it
is this difference which has been shown to significantly improve
the agreement with the observed shape of the P33 resonance~\cite{H93}.

Altogether it is a fair statement to say that the Skyrme model and
appropriate extensions work reasonably well in the $\pi N$ system
although this statement has not been analysed in terms of underlying 
form factors (except for the case of the P33 resonance in ~\cite{H93}).
It is therefore an interesting question to ask 
whether form factors extracted from the Skyrme model will
work in the $NN$ system.
It is the purpose of this paper to investigate this question.

In Sec.~II, we derive the strong
$\pi NN$ form factor in the Skyrme model, and in Sec.~III we apply
this form factor
in the NN system.
The paper is concluded in Sec.~IV.

\section{ The strong $\pi N N$ form factor}

Analysing the meson-baryon scattering S-matrix in the soliton
sectors of effective meson Lagrangians does not require to separately 
consider meson-baryon form factors: the spatial structure of the
interaction is determined by the selfconsistently calculated soliton
profiles which naturally enter in a consistent way into the scattering
equations.  This holds, 
of course, also for the analysis of the baryon-baryon interaction, or for 
the structure of the
deuteron or other nuclei. Still, there have been attempts to extract 
meson-baryon form factors from 
soliton solutions of mesonic actions, which would allow for a comparison 
with form factors
typically used in conventional meson-exchange models of the baryon-baryon 
interaction.

In a fully consistent formulation in terms of soliton and soliton
fluctuations the resulting S-matrix will not depend on the choice of
the field which interpolates between the asymptotic mesonic scattering
states. Similarly, a form factor to be used for dressing conventional
meson-baryon vertices should not depend on the choice of the
interpolating field from which it is extracted. This raises the
question whether it is possible at all to unambiguously extract form
factors from effective meson theories. In the following we will argue
that this is indeed possible if one takes due care of the local
metric associated with a given choice of interpolating field.

These metrical factors have been disregarded in early attempts to
relate the strong form factors to the soliton
profiles~\cite{Coh86,KMW87}. The procedure suggested by 
Cohen~\cite{Coh86} led to
a shape of  $G_{\pi N N}(t)$ which for small values of the momentum 
transfer $q^2$ was roughly
compatible with the conventionally used monopole form, Eq.~(1), 
but the resulting 
values of $\Lambda\approx 0.6$ GeV were less than half of the 1.3--1.7 
GeV typically required in OBE
potentials~\cite{Mac89,MHE87}. 
Later extensions including vector mesons explicitly in the 
effective action~\cite{KMW87}
led to some improvement ($\Lambda\approx0.85$ GeV) without really 
resolving the problem.

In the following, we first give the general argument how the procedure 
in Refs.~\cite{Coh86,KMW87} to
relate the strong form factors to the soliton profiles should be modified. 
We then calculate the 
$\pi N N$ form factors for a purely pionic effective action (for the
Skyrme model, and for its extension to chiral order six) 
and for the standard minimal action which includes $\rho$ and $\omega$ mesons.

The procedure followed in Refs.~\cite{Coh86,KMW87} is based on the equation of 
motion (EOM) for a pion
field $ \bf \pi $ coupled to a 
(fermionic) axial source
\be
(\Box + m_\pi^2)\pi^a(x) = J^a_5(x).
\ee
Taking matrix elements for nucleon states and using translational 
invariance leads to
\be
(-q^2+m_\pi^2) <N(p')|\pi^a(0)|N(p)> = <N(p')|J^a_5(0)|N(p)> 
\ee
with $q=p'-p$. The matrix element on the right-hand side defines the 
form factor 
$G_{\pi N N}$ through
\be
<N(p')|J^a_5(0)|N(p)> = G_{\pi N N}(-q^2)\; \bar u(p') i \gamma_5 \tau^a 
u(p)
\ee
while the matrixelement on the left-hand side to lowest order in 
$\hbar/N_c$
is the Fourier transform of the classical meson field
\be
 <N(p')|\pi^a(0)|N(p)> = \int e^{iqx} \pi^a_{cl}(x)\, dx .
\ee
Through (4),(5), and (6) the $\pi N N$ form factor thus is expressed in 
terms of the classical solution for the chiral field. 
It implies that in 
an EOM for the fluctuating pion field derived from any chiral
effective action (conveniently formulated in terms of a unitary 
matrix field 
$ U=\sigma + i\, 
\mbox{\boldmath $\tau$}
\cdot 
\mbox{\boldmath $\pi$}
$)
 \be
(\Box + m_\pi^2)\pi^a(x) = J^a_5 [ U(x)]
\ee
the matrixelements of the functional  $J^a_5 [ U(x)] $ in baryonic 
configurations may
be identified with the corresponding fermionic matrix elements of $ 
J^a_5(x) $.

It should be noted, however, that the EOM derived from some effective 
meson action is 
not immediately obtained in the form (7), because the kinetic part will 
generally contain a local metric.
Only after a field redefinition to absorb this metric into the chiral 
field the 
correspondingly transformed source function can be compared with the 
fermionic matrix
elements and the form factor. Evidently, this metric can only be 
identified from the
time-derivative part of the action, because any deviation of the spatial 
part from the 
required structure  $\nabla^2 \pi^a$ could be absorbed into the  source 
function
$ J^a_5 [ U(x)]$ without a redefinition of the field.

In terms of the Maurer-Cartan forms 
\be
L^\mu = U^\dagger \partial^\mu U = L^\mu_a \tau_a
\ee
the kinetic part ${\cal T }$ of the Lagrangian which determines the
dynamics of the field fluctuations generally is given 
by  
\be
{\cal T} =- \frac{f_\pi^2}{2} \int \;L^0_a M_{ab} L^0_b \;d^3x
\ee
with
\be
L_a^0 = 
i(-\dot{\sigma}\pi_a+\sigma\dot{\pi}_a+(
\mbox{\boldmath $\pi$}
\times
\dot{\mbox{\boldmath $\pi$}}
)_a).
\ee
This also holds for effective theories which contain more than two
time derivatives in their chiral action, because ${\cal T }$ is
obtained by expanding the Lagrangian to second order in the fluctuations.
In the Skyrme model and related models the classical field
configuration $ \pi^a_{cl}(x)$ which characterizes 
the baryon is the hedgehog
 $U_0 = \exp (i
\mbox{\boldmath $\tau$}
\cdot 
\mbox{\boldmath $x$}
F(r))$ with chiral profile
$F(r)$, rotating in isospace. 
For solitons of this type the only isovector which can 
appear in the metric tensor $M_{ab}$ is the pion field itself,
($ 
\mbox{\boldmath $ \pi $}
=|
\mbox{\boldmath $ \pi $}
|
\hat{\mbox{\boldmath $ \pi $}}
$), therefore $M_{ab}$ has to be of the form
\be
M_{ab} = M_L \hat{\pi}_a  \hat{\pi}_b + M_T (\delta_{ab}- \hat{\pi}_a  
\hat{\pi}_b)
\ee
with longitudinal and transverse metrical factors $M_L$ and $M_T$ 
depending on $\sigma$
and 
$|\mbox{\boldmath $ \pi $}|$.
The metric in (9) can be removed from the 
kinetic energy by redefining
\be
\tilde{L}^0_a = M^{1/2}_{ab} L^0_b.
\ee
For the hedgehog soliton 
$\mbox{\boldmath $ \pi $}$
rotating in isospace 
with angular
velocity $\bf{\Omega}$, the time derivative 
$\dot{\mbox{\boldmath $ \pi $}}$
is 
purely transverse while
the scalar part $\sigma $ is static
\be
\dot{\mbox{\boldmath $\pi$}}
= {\bf\Omega}\times 
\mbox{\boldmath $
\pi $}, 
\;\;\;\;\;\;\; 
\dot{\sigma}= 0.
\ee
This means that in this case $\tilde {L}^0_a $ absorbs only the 
transverse part of the metric
\be
\tilde {L}^0_a =i \sqrt{M_T} (\sigma \dot{\pi}_{a} 
+(\mbox{\boldmath $
\pi $}
\times
\dot{\mbox{\boldmath $\pi$}}
)_a)
\ee 
and we have
\be
-\tilde {L}^0_a \tilde {L}^0_a = \dot{\tilde{\pi}}_{a} 
\dot{\tilde{\pi}}_{a}
\ee
with redefined field $\tilde{\pi}_{a} = \sqrt{M_T} \pi_{a}$. This may 
seem a bit surprising
because $\bf{\pi}$ is longitudinal (by definition), but it is clearly 
a consequence of the
fact that the redefinition is determined through the time derivatives of 
the field.

Combining now Eqs. (4), (5) and (6) with $ \pi^a_{cl}(x)$ replaced by
$ \sqrt{M_T} \pi^a_{cl}$,
the $\pi N N$ form factor in the Breit frame then is obtained as 
\be
G_{\pi N N}({q}^2)= \frac{8\pi}{3}\frac{M_N f_\pi}{q}
({q}^2 + m_\pi^2) \int_0^\infty \!dr \,r^2 j_1(qr) \sqrt{M_T(r)} 
\;\sin F(r)
\ee
where $M_T(r)$ derives from the effective Lagrangian used to 
determine $F(r)$; $M_N$ and $m_\pi$ denote the nucleon and pion
masses, respectively, and $f_\pi$ is the pion decay constant.
Notice that we have changed our notation in Eq.~(16)
defining now $q\equiv |{\bf q}|$ which will be used for the remainder of
this paper.

As a typical example, we consider the standard Lagrangian for 
pseudoscalars with the dominant fourth and sixth order terms
\be
{\cal L}_{PS}={\cal L}^{(2)} + {\cal L}^{(4)} + {\cal L}^{(6)}
\ee
\be
{\cal L}^{(2)}=\frac{f_\pi^2}{4}\int (-trL_\mu L^\mu+m_\pi^2 
tr(U+U^\dagger-2))d^3x, 
\ee
\be
 {\cal L}^{(4)}=\frac{1}{32e^2}\int tr[L_\mu,L_\nu]^2 d^3x, 
\;\;\;\;\;\;\;\;\;\;\;\;
 {\cal L}^{(6)}=-\frac{1}{2}\left(\frac{3g_\omega}{m_\omega}\right)^2 
\int B_\mu B^\mu \; d^3x
\ee
and baryon current $
B_\mu=\frac{1}{24 \pi^2} \epsilon_{\mu\nu\rho\sigma} tr L^\nu L^\rho 
L^\sigma
$. It leads to the transverse metric to be used in (16)
\be
M_T(r)= 1+\frac{1}{e^2f_\pi^2}(F'^2+\frac{\sin^2 F}{r^2})+
\left(\frac{3g_\omega / m_\omega}{2 f_\pi \pi^2}\right)^2 \frac{\sin^2 
F}{r^2} F'^2.
\ee
In the original Skyrme model the term ${\cal L}^{(6)}$ is not present.
The Skyrme term $ {\cal L}^{(4)}$ therefore has to be supplied with
sufficient strength ($3.5<e<4.5$) to allow for reasonable soliton size.
In the presence of a suitable sixth-order term comparable radii can be 
obtained with reduced fourth-order strength ($6<e<7$).
Both terms may be considered as local remnants of eliminated
vector mesons.  Therefore it may be of interest to extract the 
$\pi NN$ form factor also from chiral models with explicit inclusion 
of vector mesons. Unfortunately,
there are many ways to construct such models
and for reasons of simplicity and definiteness we select a
minimal model which comprises $\rho$ and $\omega$  mesons 
together with the field $U$ in a chiral-covariant way: 
\be
{\cal L}_{V\!M}={\cal L}^{(2)}+{\cal L}^{(\rho)}+{\cal L}^{(\omega)}
\ee
with
\be
{\cal L}^{(\rho)}= \int \left(-\frac{1}{8} tr \rho_{\mu\nu} \rho^{\mu\nu} 
+\frac{m_\rho^2}{4} tr(\rho_\mu
-\frac{i}{2g}(l_\mu-r_\mu))^2 \right) d^3x, 
\ee
\be
{\cal L}^{(\omega)}=\int \left(-\frac{1}{4} \omega_{\mu\nu} 
\omega^{\mu\nu} +\frac{m_\omega^2}{2} 
\omega_\mu \omega^\mu +3 g_\omega \omega_\mu B^\mu \right) d^3x.
\ee
and $l_\mu=\xi^\dagger \partial_\mu \xi, \;r_\mu=\partial_\mu \xi 
\xi^\dagger$. 
Here $\xi$ denotes the square root of $U$
\be
\xi=U^{\frac{1}{2}}=\Sigma+i\,
\mbox{\boldmath $
\tau $}
\cdot{\bf\Pi} .
\ee
In this case, to isolate the metric for the pseudoscalars we write the 
relevant kinetic parts of ${\cal L}_{VM}$
as
\be
{\cal T} = -\int \left(\frac{f_\pi^2}{4} tr(l_0+r_0)^2+\frac{m_\rho^2}{16 
g^2} \, tr(l_0-r_0)^2 \right) d^3x.
\ee
Again, for the rotating classical hedgehog we have
\be
\dot{\Sigma}=0,\;\;\;\;\;{\bf \dot{\Pi}} = 
{\bf\Omega}\times{\bf\Pi}
\ee
and obtain
\be
{\cal T}=\frac{f_\pi^2}{2}\int\left(4\,\Sigma^2 + \frac{m_\rho^2}{f_\pi^2 
g^2} \,\Pi^2\right)
{\bf \dot{\Pi}}\cdot{\bf \dot{\Pi}} \; d^3x.
\ee
With the chiral profile $F(r)$ determined from the static minimization
of (21) we have
\be
\Sigma = \cos\frac{F}{2}, \;\;\;\;\;\;\Pi= \sin\frac{F}{2}.
\ee 
The transverse metric resulting from (27) then is
\be
M_T(r)=\left( 1 +\frac{m_\rho^2}{4 f_\pi^2 g^2} 
\,\tan^2\!\frac{F}{2}\right).
\ee
Replacing the $\omega$ mesons by the baryon current in the
lowest chiral-order
local approximation $(\omega_\mu=-3 g_\omega/m_\omega^2 B_\mu)$ 
leads to the sixth-order contribution in (19).
The elimination of the $\rho$ mesons in lowest-order local approximation
$(2g\rho_\mu=i(l_\mu-r_\mu))$ leads to the Skyrme term with $e=2 g$.
If $g$ is chosen to satisfy the KSRF relation $g^2=m_\rho^2/(8f_\pi^2)$
with vector meson mass $m_\rho=770$ MeV, i.e. $g= 2.925$, and 
$g_\omega \approx g$, both Lagrangians (17) and (21) stabilize solitons
of reasonable size. However, it has been observed~\cite{MeWa96} that after
renormalization of loop corrections the effective coupling constants
in the soliton sector favor a stronger Skyrme term ($e\approx 4$) 
and, correspondingly, a weaker sixth-order term ($g_\omega < 1$). This is
in accordance with ample past evidence, that the simple Skyrme model creates
soliton profiles which are well suited for many applications. 

In Fig.~1 we compare the form factors resulting from (16) and (20) 
for the pure Skyrme model with strong Skyrme term and no
sixth-order term ($e=3.5$, $g_\omega=0$; solid line in Fig.~1), 
and for the sixth-order extension with $e=2g=5.85$, and $g_\omega= 
3.1$ (dashed line in Fig.~1). Both cases lead to the same values for
the pion-nucleon coupling constant $G_{\pi N N}(0) = 0.99\; (2 
M_N/m_\pi)=13.5$ and the axial coupling contant $g_A=1.30$.
The same values for $g$ and $g_\omega$ we use also in (29) for the
form factor from the vector-meson model (dotted line in Fig.~1).

It is interesting to note that the
 $\pi NN$ form factor which arises from the vector meson
Lagrangian shows approximately a dipole form, Eq.~(1) with $n_\alpha=2$, 
with $\Lambda_\alpha \approx 1.5$ GeV. The $\omega$
mesons do not contribute at all to the pionic metric, because their coupling
$\omega_\mu B^\mu$ to the baryon current contains at most one time derivative 
of the pion field. The term $\; 2 \tan^2(F/2)$ in (29) is due to the
chiral invariant form of the $\rho$-$\pi$ coupling in (22) and causes the
deviation from the flat metric of ${\cal L}^{(2)}$. This results in the     
dipole form. 
The form factor derived from the corresponding local 
approximation (dashed line) shows an almost unchanged slope
for small $q^2$ but it 
suppresses higher momenta more efficiently and displays small
oscillations above $200 m_\pi^2$ which may be traced directly to the 
nonvanishing sixth-order term.

Increasing the strength of the Skyrme term, however, produces a 
qualitative change in the low-$q^2$ behaviour of the form factor: 
The soliton profile created through a strong Skyrme term causes 
the slope of the form factor near $q^2=0$ to become very small and, 
at the same time the curvature to become negative. 
This means that for small $q^2$ the effective $\pi NN$ coupling strength
stays much closer to its value at $q^2=-m_\pi^2$ than for comparable monopole
form factors. It is this feature of the Skyrme model which has been shown to 
improve the agreement of the calculated P33 phase shifts in $\pi$-N scattering
with the data over the whole $\Delta$-resonance region~\cite{H93}.
This very hard behaviour 
of the form factor for small $q^2$ is compensated by a very soft behaviour for 
$q^2 >$ 50 $m_\pi^2$ which cuts off higher momenta much more efficiently than
typical hard monopole form factors (cf.\ Fig.~2). Without the
sixth-order term ($g_\omega=0$) the form factor is monotonously
decreasing without oscillations.

\section{The two-nucleon system}

In this section, we will apply the $\pi NN$ form factor
extracted from the `simple' Skyrme model
in the NN system.
To facilitate the comparison with traditional work using
monopole (or dipole) meson-nucleon form factors,
we choose as our starting point
the OBE model of Ref.~\cite{Mac89}, which has also become known
as the Bonn-B potential~\cite{note2}.

An OBE potential is defined as the sum of
one-particle-exchange amplitudes ($V^{OBE}_\alpha$) of certain bosons
$\alpha$
with given spin, parity, mass, coupling, etc.. We use six bosons. Thus,
\begin{equation}
V
({\bf p'},{\bf p})
=\sum_{\alpha=\pi,\eta,\rho,\omega,\delta,\sigma}
V^{OBE}_{\alpha} 
({\bf p'},{\bf p})
[F_\alpha
(({\bf p'}-{\bf p})^2)]^2
\end{equation}
with $\pi$ and $\eta$ pseudoscalar,
$\sigma$ and $\delta$ scalar, and
$\rho$ and $\omega$ vector bosons.
Each vertex is multiplied with 
a form factor $F_\alpha$
(i.~e., two factors per OBE diagram).

For the unitarizing scattering equation, we use the
the relativistic three-dimensional
reduction of the Bethe-Salpeter equation suggested by
Blankenbecler and Sugar~\cite{BS66}:
\begin{equation}
\hat{T}
({\bf p'},{\bf p})
=\hat{V}({\bf p'},{\bf p})+
\int d^3k\:
\hat{V}({\bf p}',{\bf k})\:
\frac{M_N}
{{\bf p}^{2}-{\bf k}^{2}+i\epsilon}\:
\hat{T}({\bf k},{\bf p})
\end{equation}
where $\hat{T}$ denotes the $T$-matrix, and
${\bf p}$,
${\bf k}$, and
${\bf p}'$ are the initial, intermediate, and final relative
three-momenta, respectively, of the two interacting nucleons.
The relationship between $\hat{V}$
and $V$, the 
amplitude of Eq.~(30),
is 
\begin{equation}
\hat{V}({\bf p'},{\bf p})
 = \sqrt{\frac{M_N}{E_{p'}}}\:  V({\bf p'},{\bf p})\:
 \sqrt{\frac{M_N}{E_{p}}},
\end{equation}
with $E_{p}\equiv \sqrt{M_N^2+{\bf p}^2}$
and $E_{p'}$ similarly.
For further details 
see appendix A of Ref.~\cite{Mac89} and Ref.~\cite{Mac93}.

The meson parameters used in the original Bonn-B potential are listed
in Table I, column Bonn-B. The phase-shift predictions by Bonn-B
for neutron-proton ($np$)
 scattering below 300 MeV lab.\ energy 
are shown in Fig.~3 by the dotted lines.

In the Bonn-B model,
we replace now the monopole form factor applied to the
$\pi NN$ vertex by the `simple'
Skyrme model $\pi NN$ form factor, i.~e., Eqs.~(16) and (20) with 
$e=3.5$ and $g_\omega=0$ (solid curve in Fig.~2). 
The form factors of mesons other than the pion are not changed.

We make some minor adjustments of the coupling constants of the vector
mesons to optimize
the fit of the $P$-wave phase shifts, and we fine-tune
the coupling constant of the sigma boson to accurately fit
the $S$-wave effective range parameters and the deuteron binding
energy. The new meson parameters are listed in Table~I, column
Skyrme FF. 
The phase-shift predictions for $np$ scattering are plotted
in Fig.~3 by the solid lines
and deuteron properties are given in Table~II.
It is clearly seen that the model using the Skyrme form factor (FF)
at the pion vertex reproduces the two-nucleon data as well
as the original Bonn-B potential.

For comparison, we also show the results obtained when applying
a soft monopole form factor (with 
$\Lambda_\pi=0.8$ 
GeV)
for the pion; see dashed line
in Fig.~3.
Note that, as customary in OBE models, the sigma-boson parameters
are adjusted such as to fit the $S$-waves.
Obviously, a soft pion form factor of monopole shape yields
disastrous results for several partial-waves of NN scattering.
In particular, the mixing parameters, $\epsilon_1$ and $\epsilon_2$,
which depend entirely on the nuclear tensor force, are described
badly.
The same is true for the deuteron, see column `$\Lambda_\pi=0.8$'
of Table~II.
The common reason for all these formidable predictions is that the
soft monopole cuts out also part of the long-range tensor force created
by the pion.

It is interesting to note that, 
at large momentum transfer
($q^2\gea 80 m_\pi^2$), 
the Skyrme FF 
(solid line in Fig.~2)
is even softer than
the soft monopole form factor
(dashed line in Fig.~2).
Thus, strong suppression
at high momentum transfer does not cause problems and is,
in fact, the desired property of a form factor.

On the other hand, 
at low $q^2$, 
the Skyrme FF
stays close to its value at the meson
pole and at $q^2\approx 0$.
In contrast, the soft monopole 
falls off drastically already at low $q^2$.
This causes problems in the NN system since it modifies
the long-range part of the nuclear force.
It also contradicts the idea of a form factor which is 
to regularize the short-range interaction.

For many years, it has been a great puzzle why NN models
{\it seemingly}
need a very hard $\pi NN$ form factor.
Based upon the above discussion,
one can now explain this.
Traditionally, OBE models use form factors of monopole shape
which have the undesirable feature of cutting down also the 
low-$q^2$ region.
The only way to avoid this within the monopole concept is
to use a very large cutoff mass, like $\Lambda_\pi = 1.7$ GeV
in the Bonn-B potential (cf.\ dotted curve in Fig.~2).
This large cutoff mass then suggests that the required form factor
is very hard. However, this is misleading.
The large cutoff mass is needed to avoid an unreasonable suppression
of the low-$q^2$ (equivalent to long-range) region.
If this unwanted low-$q^2$ suppression can be avoided, a
soft form factor is no problem in the NN system.
The Skyrme FF proves the point.

There is one last item that deserves attention.
The Bonn-B potential uses for the $\pi NN$ coupling constant
the large value $g^2_\pi/4\pi=14.4$.
In models that apply a monopole for the pion,
a large value for the $\pi NN$ coupling constant is
needed to predict the deuteron quadrupole moment correctly.
However, recent determinations of the $\pi NN$ coupling
constant have yielded the value $g^2_\pi/4\pi=13.5\pm 0.1$
\cite{STS93} which is substantially smaller than the one above.
As discussed in Refs.~\cite{MS91,ML93},
the deuteron quadrupole moment is predicted far too small with
$g^2_\pi/4\pi=13.5$ in OBE models using monopole form factors.

An important by-product of our present investigation is 
the result that
there is no such problem when the
Skyrme FF is used.
We use $g^2_\pi/4\pi=13.5$ when applying the Skyrme FF for the pion,
and the deuteron quadrupole moment, $Q_d$, is then predicted to be
0.274 fm$^2$ which is within the  empirical range (cf.\ Table~II).
Note that, applying a monopole with $\Lambda_\pi = 1.7$ GeV, 
$Q_d=0.266$ fm$^2$ is predicted
when $g^2_\pi/4\pi=13.5$ is used~\cite{ML93}.
The deuteron quadrupole moment is a long-range property and, thus,
sensitive to the low-$q^2$ behaviour of the form factor.
Again, the large values of the Skyrme FF at low $q^2$ 
are clearly preferred by the NN system.

\section{Conclusions}

We have shown in this paper how to extract meson-baryon form factors
from the soliton sector of effective meson theories which do not 
depend on the choice of the field that interpolates between the
asymptotic meson states. The crucial ingredient is a redefinition of
this field to absorb the local metric which characterizes the kinetic
energy of the fluctuating field. The axial source in the resulting
flat metric then can be used to extract the form factor in the usual
way. 

We have applied this procedure to two standard examples of effective
meson theories: The minimal chiral model for $\pi , \rho $ and
$\omega$ mesons, and the Skyrme model (with or without sixth-order
extension). Both models work qualitatively well in the $\pi N$ system
at least to the extent we could expect from one- or two-parameter
models. 

The resulting strong form factors are considerably affected by the 
respective local metric. Previous attempts~\cite{Coh86,KMW87}
in which the metrical factors were omitted had led to very soft
form factors of the conventional monopole type for low $q^2$.
Our result for the chiral $\pi \rho \omega$ model is close to a dipole
form with a cutoff mass of about 1.5 GeV. This difference, however,
(which originates in the chiral covariant $\rho \pi \pi$ coupling)
is not sufficient for substantial improvement in the application
of OBE potentials to the two-nucleon system.

On the other hand, the Skyrme term is responsible for a qualitative
change in the form factor:
It starts with almost vanishing slope and negative
curvature for low $q^2$, and then falls off much faster than
comparable monopole form factors. In OBE potentials this very hard
behaviour for low $q^2$ provides the necessary strength for the tensor
force while at the same time the high momenta are still efficiently
cut off. It is remarkable that in order to have the full advantage of
this effect it is necessary to employ a Skyrme term with sufficient
strength ( Skyrme parameter $e\approx4$, or less). The magnitude of 
$e$ which derives from the elimination of $\rho$ mesons ( $e=2g\approx
6-7$) is not sufficient. The fact that $e\gea 2g$ does not lead to a 
satisfactory soliton has been noticed in many instances and is
supported by the recent discussion of loop corrections in the soliton 
sector.

We have applied the $\pi NN$ form factor based upon the strong
Skyrme term
in the two-nucleon system using the OBE model
for the NN interaction.
Deuteron properties and phase parameters of NN scattering are reproduced
well. 

Traditional OBE models use form factors of monopole shape and 
require a very hard pion form factor.
This has been a long-standing puzzle.
A comparison of the soft monopole with the Skyrme FF
reveals that
the latter leaves low momentum transfers essentially
unaffected while the former also suppresses the
low-momentum region.
To avoid the low-$q^2$ suppression, the monopole needs a large
cutoff-mass parameter which results in an over-all hard form 
factor.

Because of its strong suppression of large momenta,
the Skrme FF can be termed as soft. 
On the other hand, since it does not suppress low momenta,
it is compatible with the NN system.
Deuteron properties can be reproduced with the small $\pi NN$
coupling constant $g^2_\pi/4\pi=13.5$, which does not work with
the monopole.

In summary, the Skyrme FF is a soft pion form factor that is
compatible with the $\pi N$ and $NN$ system.
This is impossible to achieve with form factors of
monopole shape.

\vspace*{1cm}
This work was supported in part by the U.S.\ National Science Foundation
under Grant No.\ PHY-9211607.

\pagebreak

\begin{figure}
\caption{Form factors emerging from Skyrme-type models.
The solid line results from the `simple' Skyrme model
(Eqs.~(16) and (20) with $e=3.5$ and $g_\omega=0$),
while the dashed line includes a sixth-order term
($e=5.85$ and  $g_\omega=3.1$).
The dotted line is based upon the vector meson model Eq.~(21).
All form factors are normalized to unity at the pion pole.}
\end{figure}

\begin{figure}
\caption{Comparison of different $\pi NN$ form factors. The solid line
represents the form factor extracted from the `simple' Skyrme model 
(same as solid line in Fig.~1).
The dashed and dotted lines are monopole form factors [Eq.~(1)
with $n_\alpha=1$]
with cutoff masses $\Lambda_\pi = 0.8$ and 1.7 GeV, respectively.}
\end{figure}

\begin{figure}
\caption{Neutron-proton ($np$) phase-shifts, $\delta$, 
and mixing parameters, $\epsilon$, 
for $J\leq 2$ below 300 MeV laboratory energy, $T_{lab}$.
The solid lines show the predictions by the present model using
the Skyrme $\pi NN$ form factor.
The dotted lines are the predictions by
the original Bonn-B model which applies a monopole form factor
with $\Lambda_\pi=1.7$ GeV
at the $\pi NN$ vertex, while the dashed lines 
are obtained by applying a
monopole with $\Lambda_\pi=0.8$ GeV for the pion.
Open circles represent the 
Nijmegen multi-energy $np$ phase shift analysis~[22],
and solid dots are from the VPI single-energy analysis
VS35~[23].}
\end{figure}

\pagebreak

\begin{table}
\caption{Meson parameters used in the OBE potential models
considered in the present work.}
\begin{tabular}{llllllll}
  &  &  &  & \multicolumn{3}{c}{{\bf Bonn-B}$^a$}   & {{\bf Skyrme FF}$^b$} \\
meson & $J^P$ & $I$ & $m_{\alpha}$ (MeV)  
     &  $g^{2}_{\alpha}/4\pi$
            [$f_\alpha/g_\alpha$] 
     &  $\Lambda_{\alpha}$ (GeV)
     &  $n_\alpha$
     &  $g^{2}_{\alpha}/4\pi$ 
             [$f_\alpha/g_\alpha$] 
\\
\hline 

$\pi$ & $0^-$ & 1 & 138.03  & 14.4 & 1.7&1 & 13.5  \\

$\eta$ & $0^-$ & 0 & 548.8  &   3  & 1.5&1 & 3  \\

$\rho$ & $1^-$ & 1 & 769    &  0.9 [6.1]  & 1.85&2 & 0.9 [6.3] \\

$\omega$ & $1^-$ & 0 & 782.6 & 24.5   &  1.85&2 & 26  \\

$\delta$ & $0^+$ & 1 & 983  & 2.488  &   2.0&1 & 2.488 \\

$\sigma^{c}$ & $0^+$ & 0 & 550 & 8.9437  &  1.9&1 & 9.4369 \\
             &       &   & (720) &(18.3773) &(2.0)&(1)&(19.5806)\\
\end{tabular}
\small
$^a$ Ref.~\cite{note2}; for definition of $\Lambda_\alpha$ and
$n_\alpha$ see Eq.~(1).
\\
$^b$ OBE model that uses the `simple' Skyrme model form factor
for the pion; see text for details.
\\
$^c$ The $\sigma$ parameters given in parenthesis
 apply to the $T=0$
 $NN$ potential,
while the unparenthesized values are for $T=1$.
\end{table}

\begin{table}
\caption{
Deuteron properties
as predicted
by OBE potential models discussed in the text
and from experiment.}
\begin{tabular}{lllll}
     & Bonn-B & Skyrme FF & $\Lambda_\pi=0.8^a$ & Experiment \\
\hline  
Binding energy (MeV) & 2.2246  & 2.22454 &2.2246& 
2.224575(9)$^b$  \\
D-state probability (\%) & 4.99 &  4.71 &2.54& --- \\
Quadrupole moment (fm$^{2}$) & 0.278& 0.274&0.242& 0.276(3)$^c$\\
Asymptotic D/S-state &   0.0264 & 0.0257 &0.0236&   0.0256(4)$^d$ \\
\end{tabular}
\small
$^a$ OBE model that uses a monopole form factor with $\Lambda_\pi=0.8$
GeV for the pion.
\\
$^{b}$ Ref.~\cite{LA82}.
\\
$^{c}$ 
Corrected for meson-exchange currents and relativity~\cite{note3}.
\\
$^d$ Ref.~\cite{RK90}.
\end{table}


\begin{thebibliography}{99}
\bibitem{Mac89} R. Machleidt, Adv. Nucl. Phys. {\bf 19}, 189 (1989).
\bibitem{NRS78} M. M. Nagels, T. A. Rijken, and J. J. de Swart,
Phys. Rev. D {\bf 17}, 768 (1978).
\bibitem{Lac80} M. Lacombe {\it et al.}, Phys. Rec. C {\bf 21},
861 (1980).
\bibitem{MHE87} R. Machleidt, K. Holinde, and C. Elster, Phys. Reports
{\bf 149}, 1 (1987).
\bibitem{ML94} For a recent critical review, see: 
R. Machleidt and G. Q. Li, Phys. Reports
{\bf 242}, 5 (1994).
\bibitem{Tho83} A. W. Thomas, Adv. Nucl. Phys. {\bf 13}, 1 (1983);
G. A. Miller, Intern. Rev. Nucl. Phys. {\bf 1}, 190 (1984).
\bibitem{note1} See Fig.~4.6 (p.~233) and related text of Ref.~\cite{Mac89}.
\bibitem{PDG94} Particle Data Group, Phys. Rev. D {\bf 50}, 1173 (1994).
\bibitem{HT90} K. Holinde and A. W. Thomas, Phys. Rev. C {\bf 42}, 1195
(1990); 
T. Ueda, Phys. Rev. Lett. {\bf 68}, 142 (1992);
G. Janssen, K. Holinde, and J. Speth, Phys. Rev. Lett. {\bf 73}, 1332 (1994).
\bibitem{TTM81} A. W. Thomas, S. Th\'eberge, and G. A. Miller, 
Phys. Rev. D {\bf 22}, 2838 (1980).
\bibitem{HEH84} A. Hayashi, G. Eckart, G. Holzwarth, and H. Walliser,
       Phys. Lett. {\bf 147B}, 5 (1984);
      M.P. Mattis and M. Karliner, Phys. Rev. D {\bf 31},
      2833 (1985); M. Karliner and M.P. Mattis, Phys. Rev. D {\bf 34},
      1991 (1986);
      G. Eckart, A. Hayashi, and G. Holzwarth, Nucl. Phys. {\bf A448}, 732
      (1986).
\bibitem{WMW92} H. Walliser, Nucl. Phys. {\bf A524}, 706 (1991);
 D. Masak, H. Walliser, and G. Holzwarth, Nucl. Phys. {\bf A536},
 583 (1992).
\bibitem{HPJ90} G. Holzwarth, G. Pari, and B.K. Jennings, 
       Nucl. Phys. {\bf A515}, 665 (1990).
\bibitem{H93}
G. Holzwarth, in: 'Baryons as Skyrme solitons',  ed.
G. Holzwarth (World Scientific, Singapore, 1993) p.~279.
\bibitem{SWHH89}B. Schwesinger, H. Weigel, G. Holzwarth, and H. Hayashi,
Phys. Rep. {\bf 173}, 173 (1989).
\bibitem{Coh86} T. D. Cohen, Phys. Rev. D {\bf 34}, 2187 (1986).
\bibitem{KMW87} N. Kaiser, U. G. Meissner, W. Weise, Phys. Lett. 
{\bf B198}, 319 (1987);\\
N. Kaiser, U. Vogl, W. Weise, U. G. Meissner, Nucl. Phys.
{\bf A484}, 593 (1988). 
\bibitem{MeWa96} F. Meier, and H. Walliser, hep-ph/9602359, Phys.
Reports (1996) (in press).
\bibitem{note2} The Bonn-B potential is presented in section 4 
and appendix A (Table A.1, Potential B) of
Ref.~\cite{Mac89}.
\bibitem{BS66} R. Blankenbecler and R. Sugar, Phys. Rev.
{\bf 142}, 1051 (1966).
\bibitem{Mac93} R. Machleidt, in: Computational Nuclear 
Physics 2---Nuclear Reactions, K. Langanke, J. A. Maruhn, and 
S. E. Koonin, eds. (Springer, New York, 1993), Chapter 1.
\bibitem{Sto93} V. G. J. Stoks, R. A. M. Klomp, M. C. M. Rentmeester,
and J. J. de Swart, Phys. Rev. C {\bf 48}, 792 (1993).
\bibitem{Arn91} R. A. Arndt, Virginia Polytechnic Institute (VPI)
and State University, Interactive Dial-In Program SAID,
Solution VS35.
\bibitem{LA82} C. van der Leun and C. Alderliesten, Nucl. Phys.
{\bf A380}, 261 (1982).
\bibitem{note3}
Note that our predictions for the deuteron quadrupole moment,
$Q_d$, are based on the 
nonrelativistic impulse approximation
and do not include meson-current and relativistic
corrections.  Therefore,
to make the comparison with the experimental data meaningful, we have
subtracted from the experimental value for $Q_d$ [0.2859(3) fm$^2$ \cite{RV75}]
the meson-exchange current and relativistic contributions,
which are 0.010 fm$^2$ for the Bonn potential
according to the most recent and very thorough calculation
by Henning~\cite{Hen93}. Thus, we list 0.276(3) fm$^2$ in the last column
of Table~II as the empirical quadrupole moment where the assigned error
of 0.003 fm$^2$ is the uncertainty which we assume for the evaluation of the
theoretical corrections. 
\bibitem{RV75} R. V. Reid and M. L. Vaida, Phys. Rev. Lett.
{\bf 34}, 1064 (1975); D. M. Bishop and L. M. Cheung, Phys. Rev. A {\bf 20},
381 (1979); T. E. O. Ericson and M. Rosa-Clot, Nucl. Phys. {\bf A405}, 497
(1983).
\bibitem{Hen93} H. Henning, private communication.
\bibitem{RK90} N. L. Rodning and L. D. Knutsen, Phys. Rev. C
{\bf 41}, 898 (1990).
\bibitem{STS93} V. Stoks, R. Timmermans, and J. J. de Swart,
Phys. Rev. C {\bf 47}, 512 (1993).
\bibitem{MS91} R. Machleidt and F. Sammarruca, Phys. Rev. Lett.
{\bf 66}, 564 (1991).
\bibitem{ML93} R. Machleidt and G. Q. Li, $\pi N$-Newsletter
{\bf 9}, 37 (1993).
\end{thebibliography}
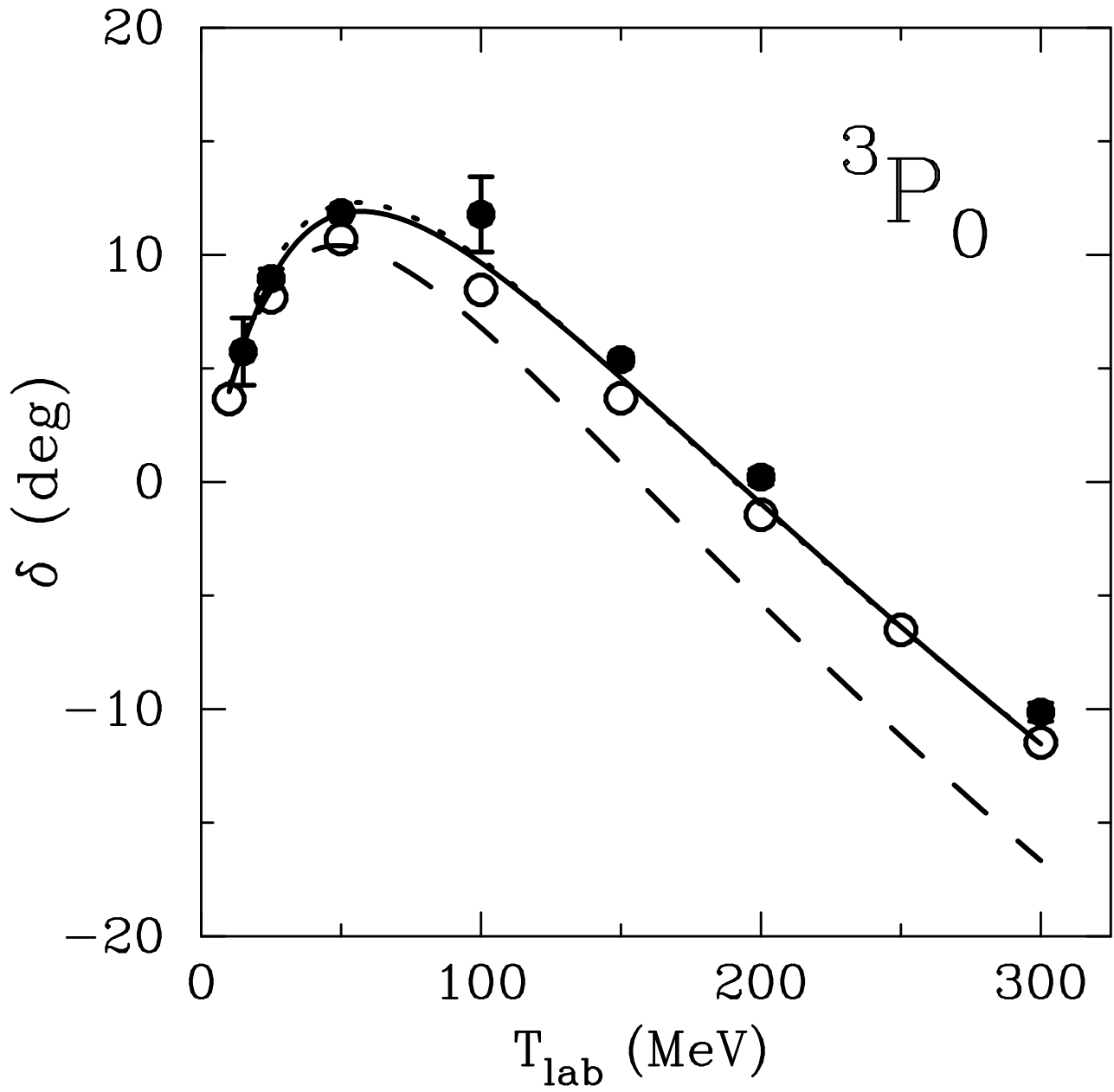
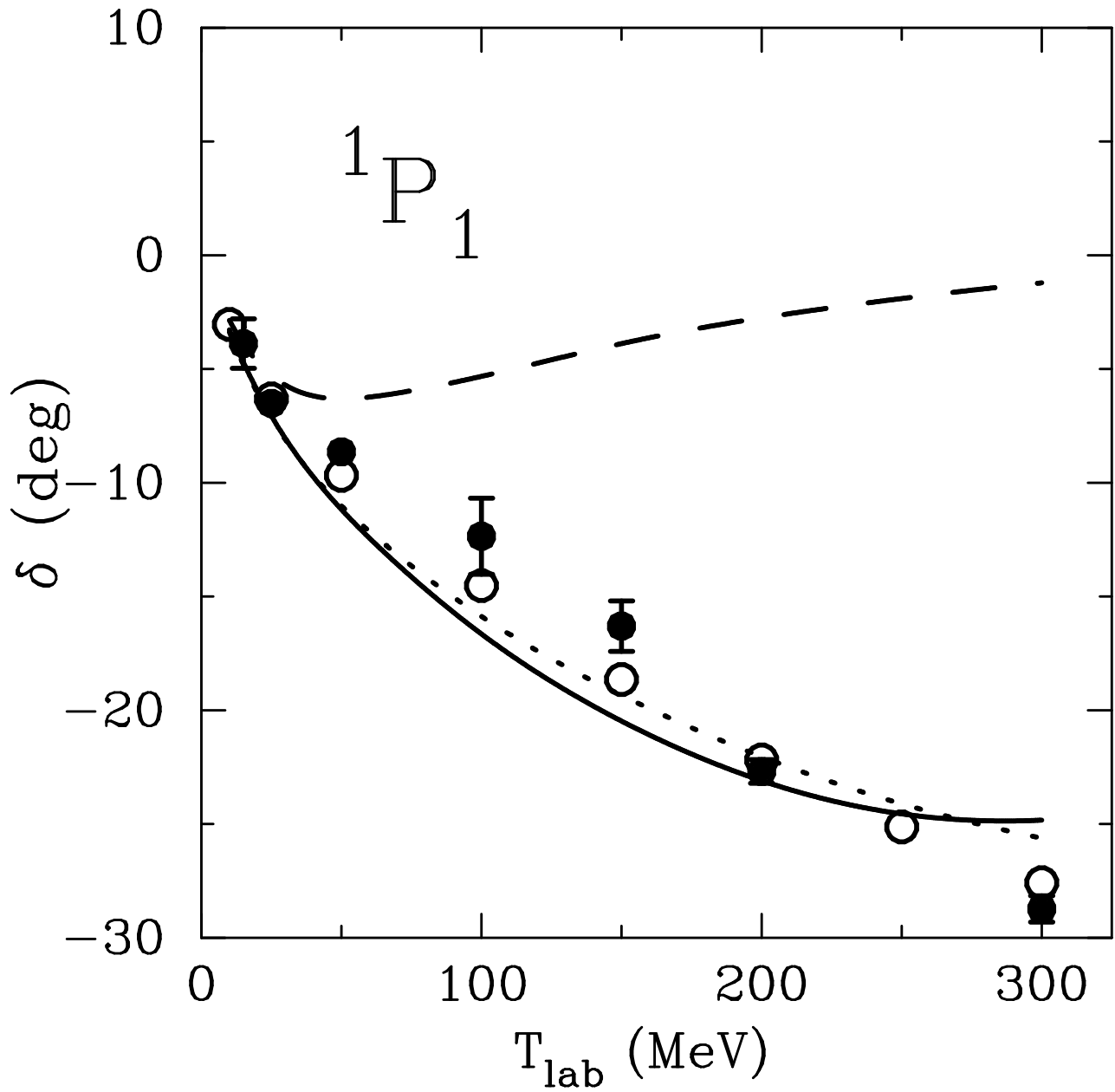
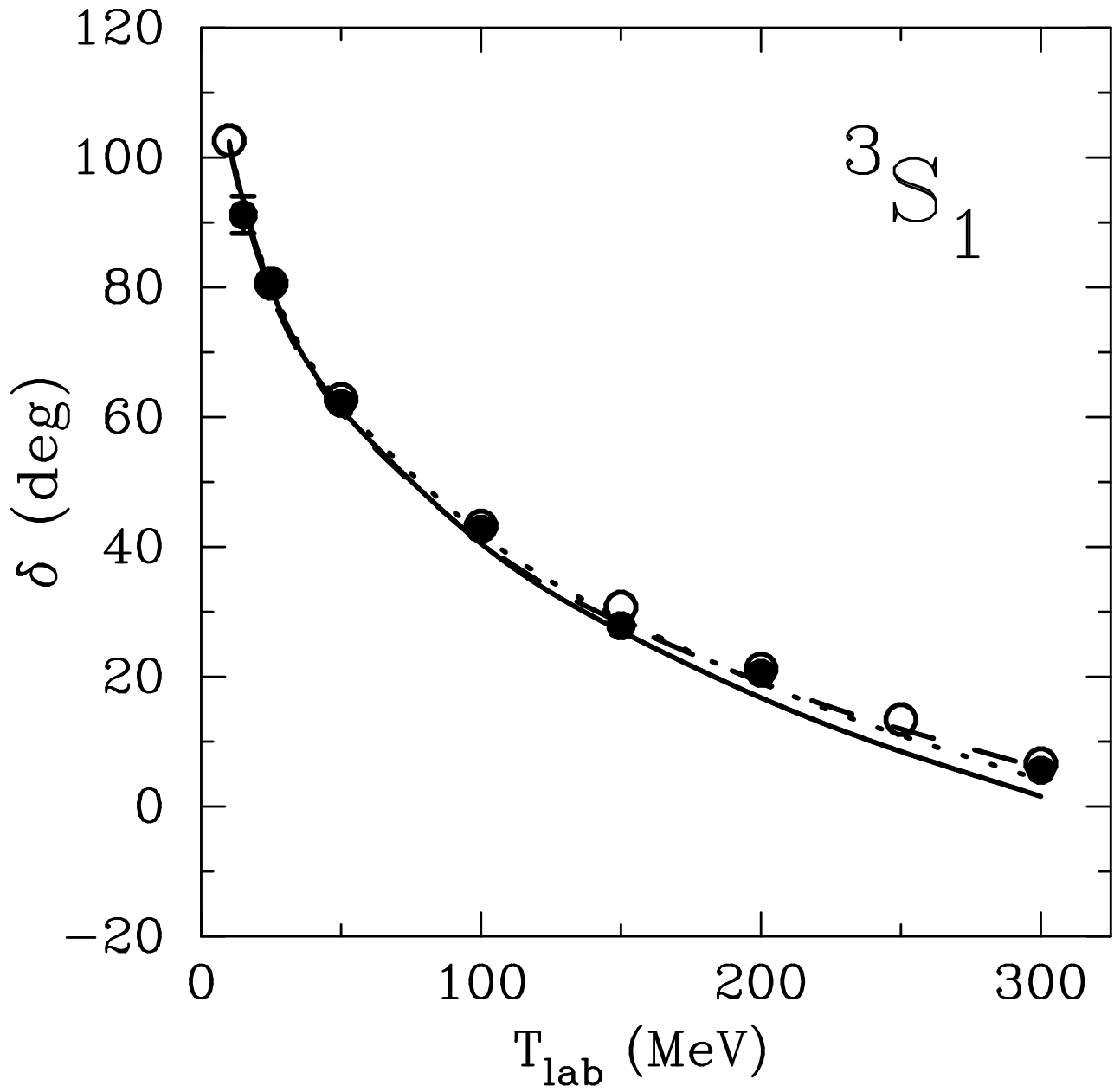
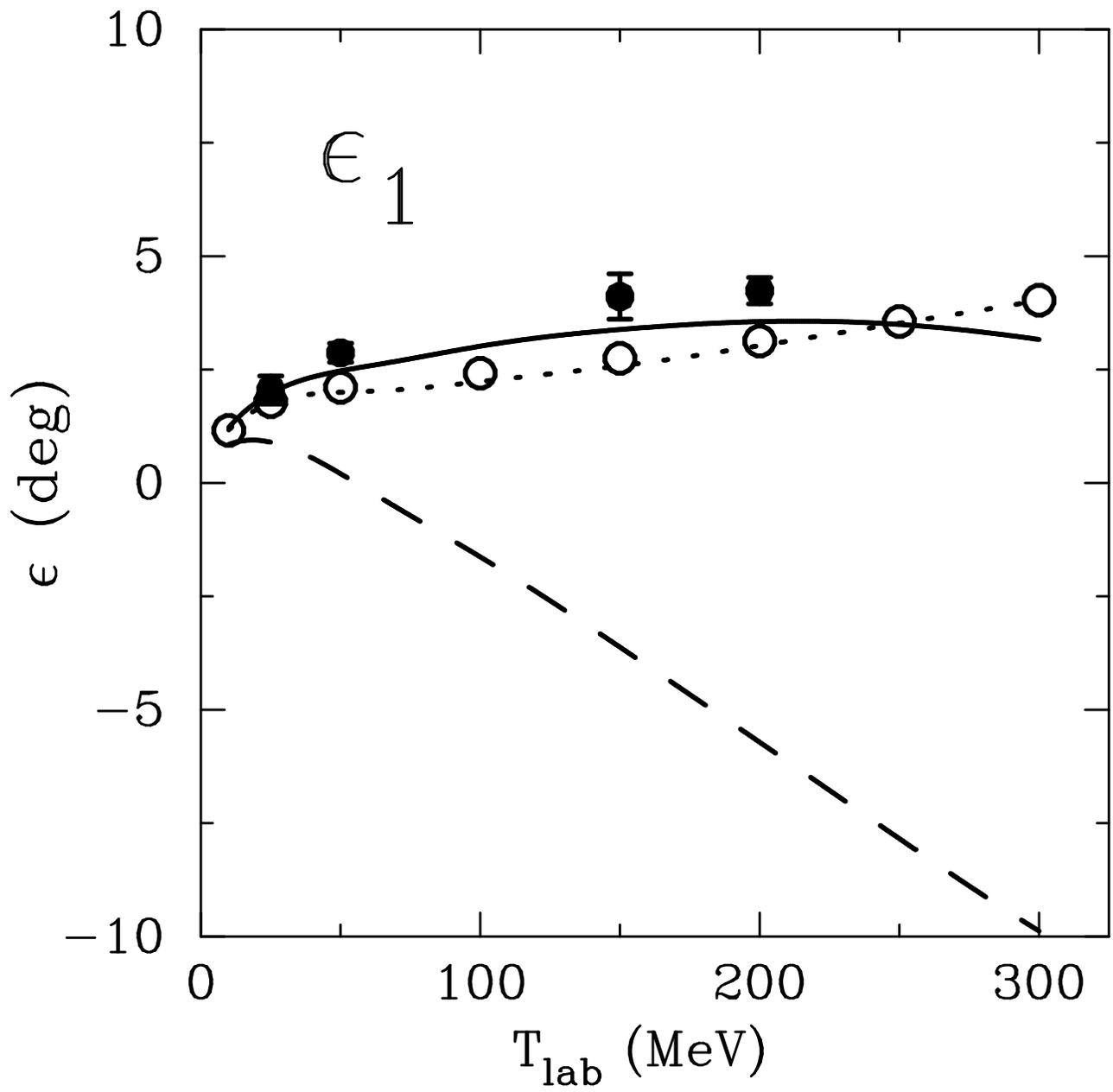
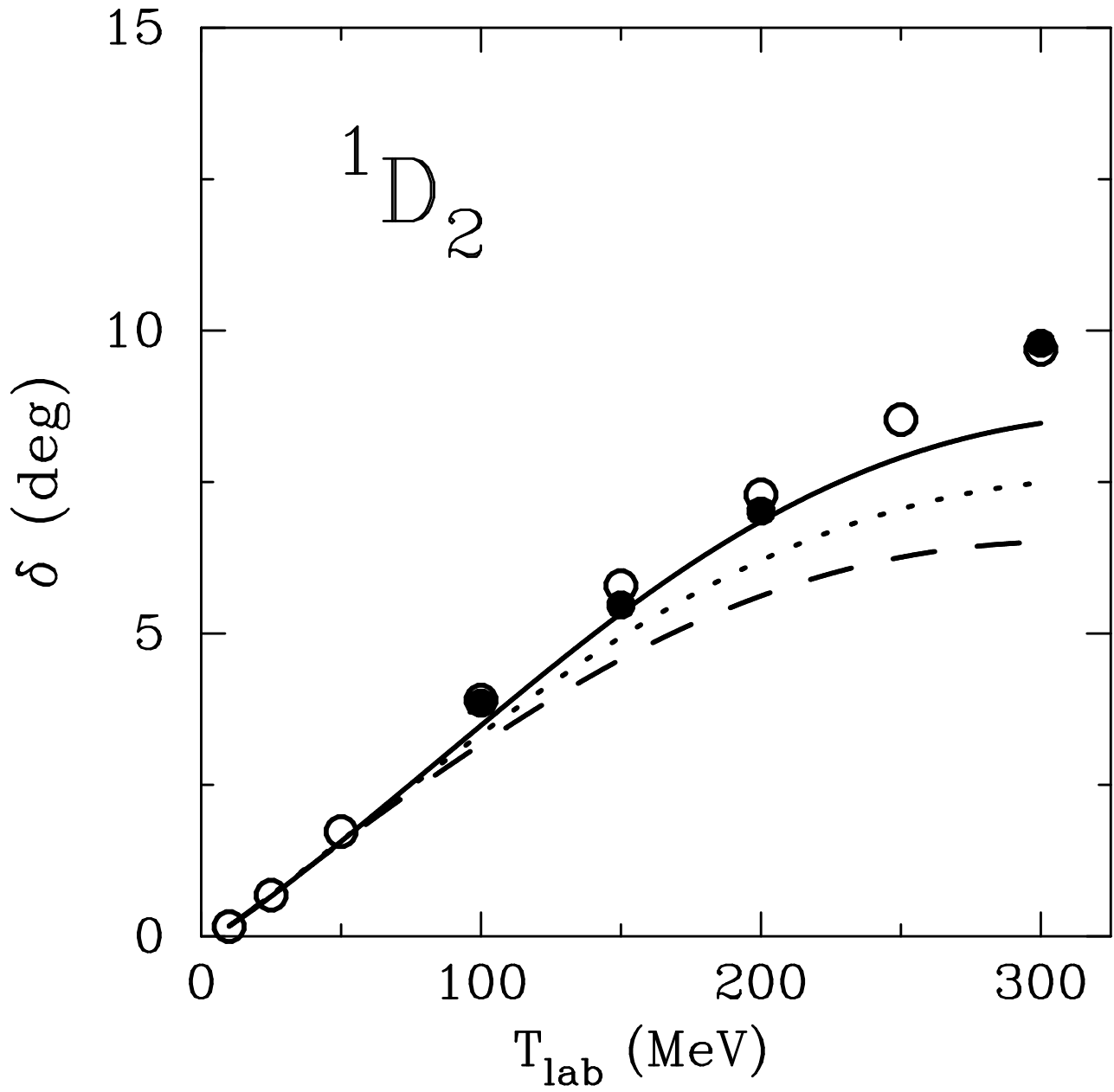
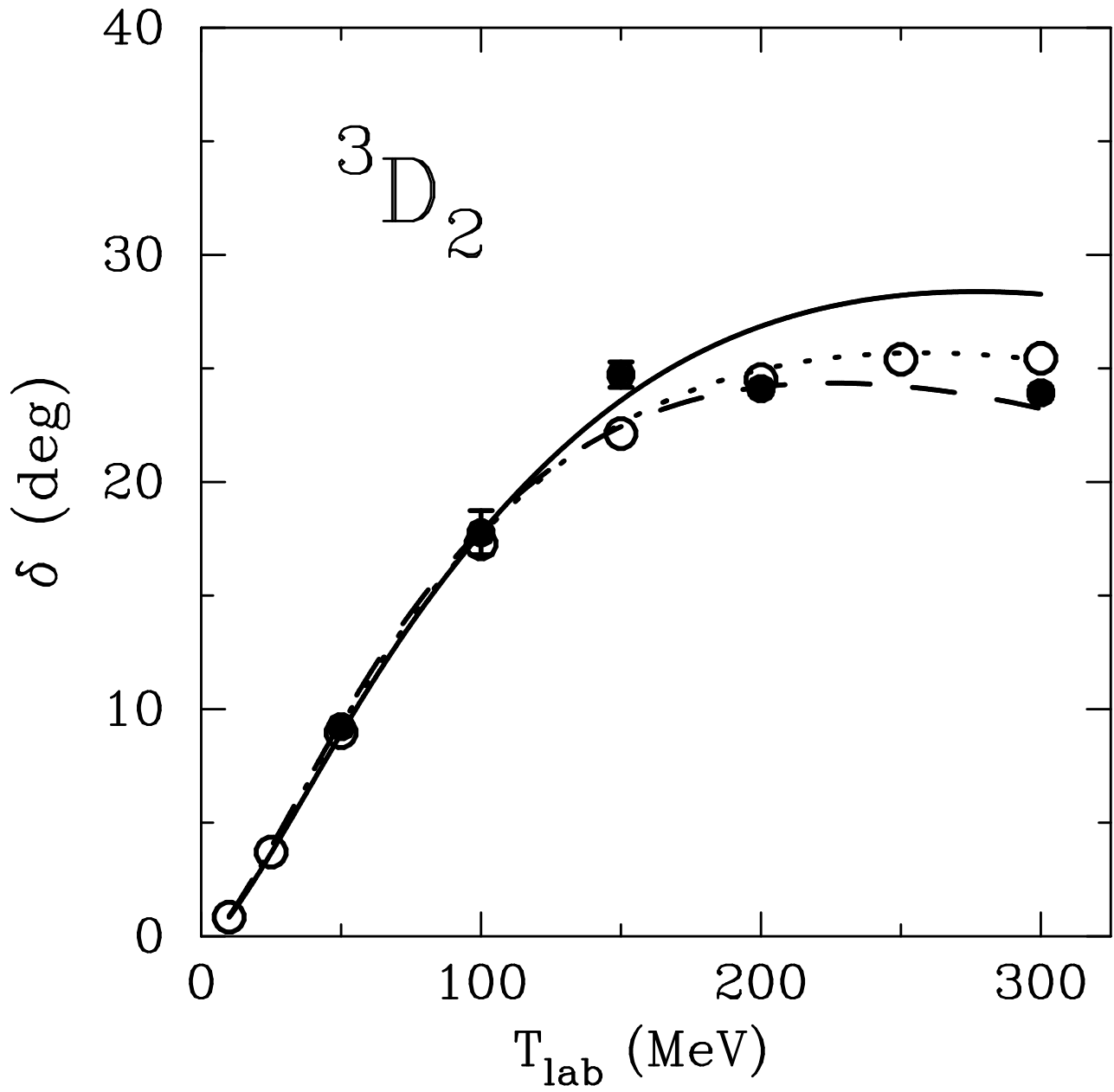
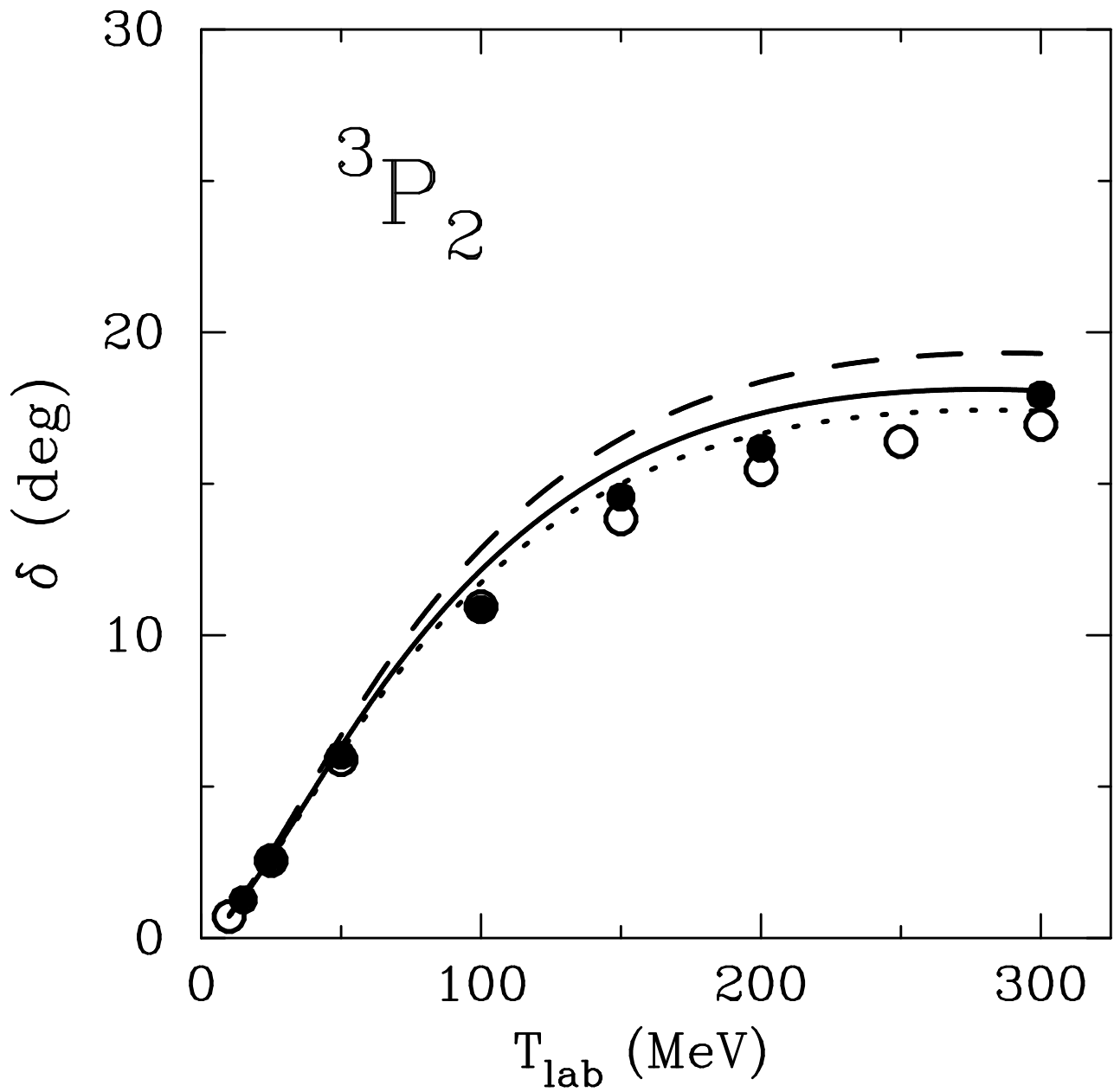
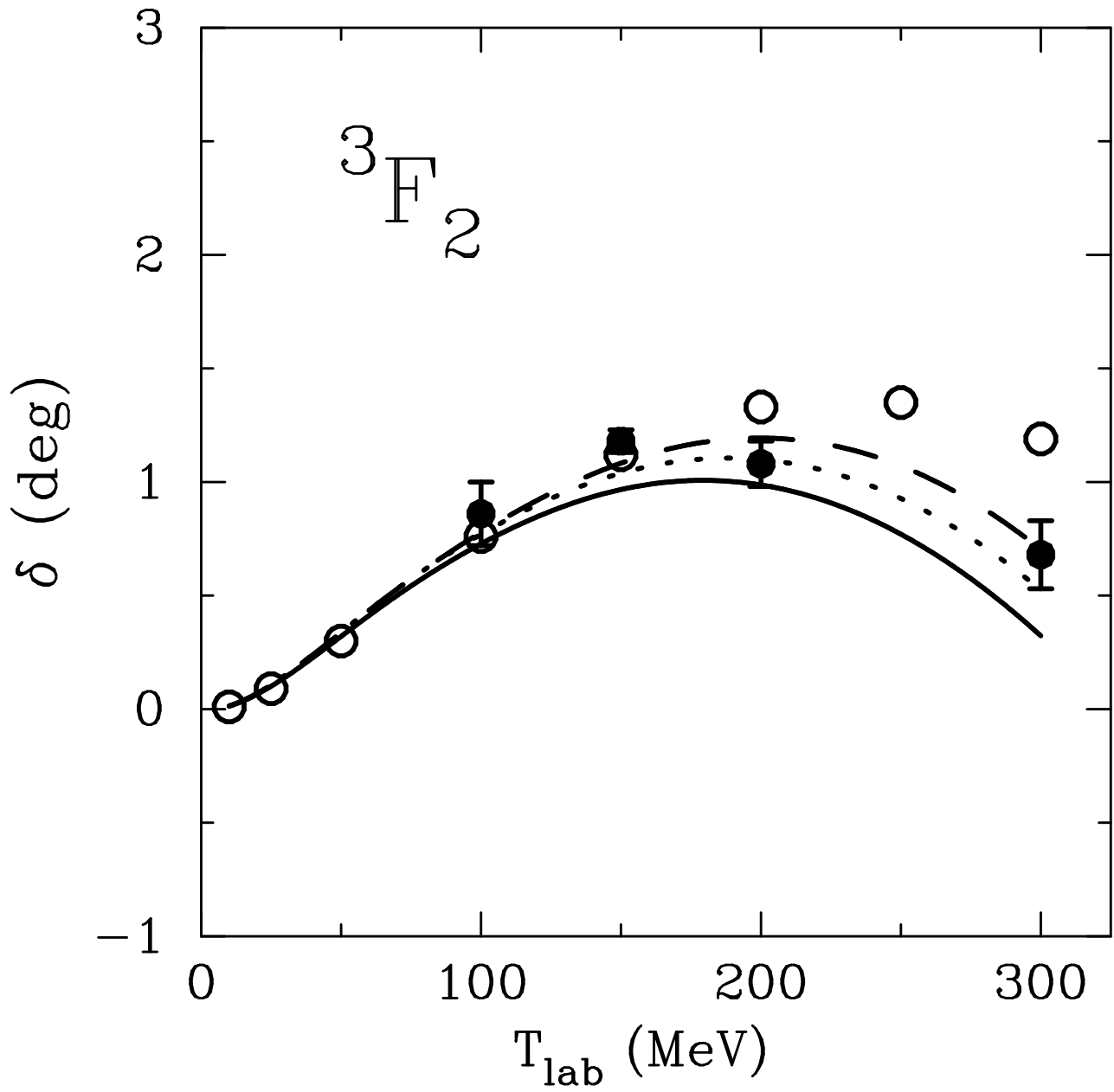
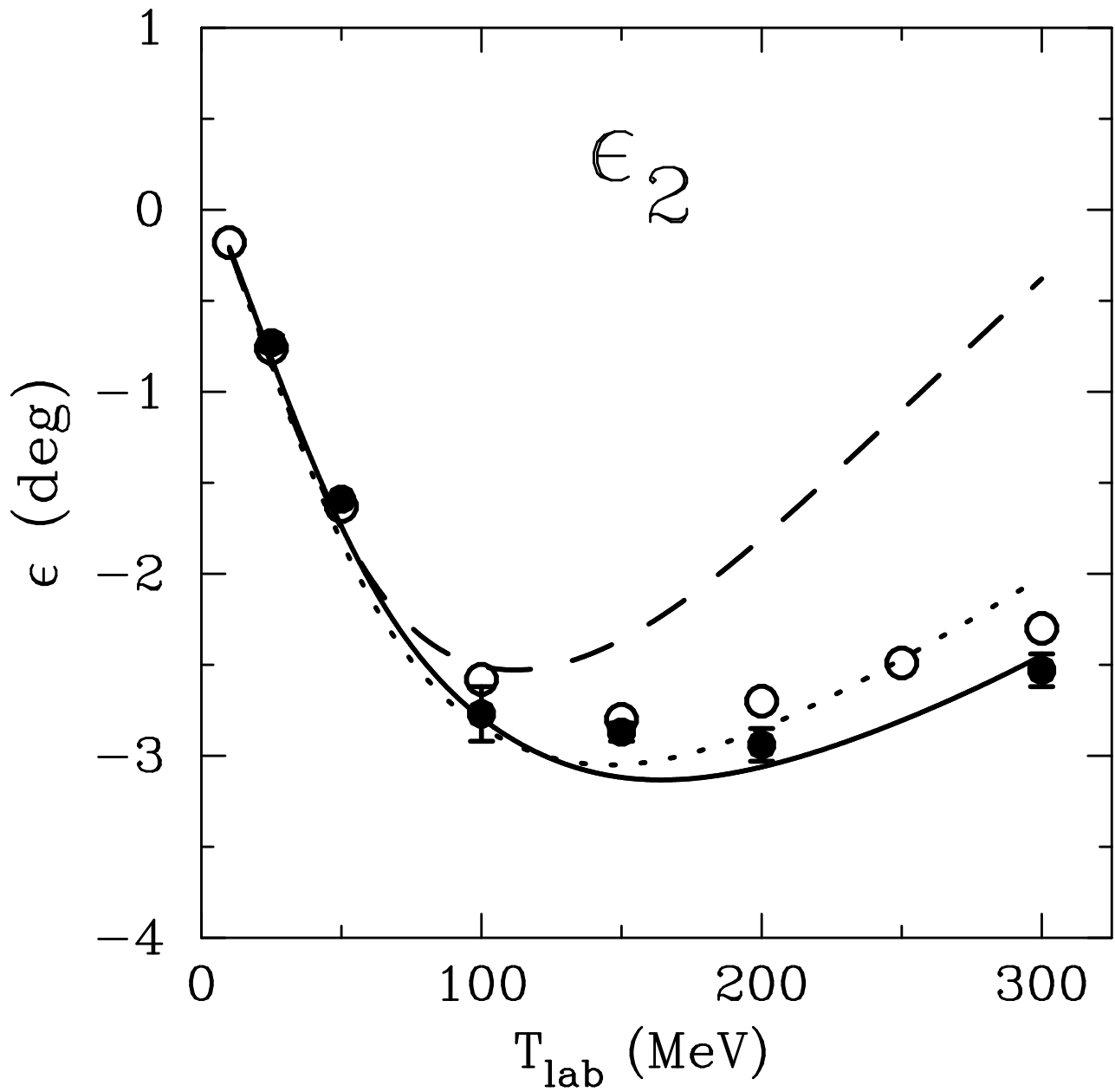
\end{document}